\documentclass[sigconf]{acmart}

\usepackage{xspace} 
\usepackage{pifont}
\usepackage{hyperref}
\usepackage{enumitem}
\usepackage{multirow}
\usepackage{tabularx}

\copyrightyear{2022}
\acmYear{2022}
\setcopyright{acmlicensed}
\acmConference[SIGIR '22]{Proceedings of the 45th International ACM SIGIR Conference on Research and Development in Information Retrieval}{July 11--15, 2022}{Madrid, Spain.}
\acmBooktitle{Proceedings of the 45th International ACM SIGIR Conference on Research and Development in Information Retrieval (SIGIR '22), July 11--15, 2022, Madrid, Spain}
\acmPrice{15.00}
\acmISBN{978-1-4503-8732-3/22/07}
\acmDOI{10.1145/3477495.3531822}

\begin{document}
\fancyhead{}
\title{How does Feedback Signal Quality Impact Effectiveness of Pseudo Relevance Feedback for Passage Retrieval?}

\author{Hang Li}
\affiliation{%
  \institution{The University of Queensland}
  \city{Brisbane}
  \country{Australia}
}
\email{hang.li@uq.edu.au}
\author{Ahmed Mourad}
\affiliation{%
  \institution{The University of Queensland}
  \city{Brisbane}
  \country{Australia}
}
\email{a.mourad@uq.edu.au}
\author{Bevan Koopman}
\affiliation{%
  \institution{CSIRO}
  \city{Brisbane}
  \country{Australia}
}
\email{bevan.koopman@csiro.au}
\author{Guido Zuccon}
\affiliation{%
  \institution{The University of Queensland}
  \city{Brisbane}
  \country{Australia}
}
\email{g.zuccon@uq.edu.au}
\settopmatter{authorsperrow=4}




\renewcommand{\shortauthors}{Trovato and Tobin, et al.}
\newcommand\todo[1]{{\color{red}#1}}
\newcommand{\COt}{CO$_2$\xspace}
\newcommand{\COtt}{CO$_2$e\xspace}
\newcommand{\COte}{gCO$_2$e/kWh\xspace}
\newcommand{\tick}{\ding{51}}
\newcommand{\cross}{\ding{55}}
\begin{abstract}

Pseudo-Relevance Feedback (PRF) assumes that the top results retrieved by a first-stage ranker are relevant to the original query and uses them to improve the query representation for a second round of retrieval. This assumption however is often not correct: some or even all of the feedback documents may be irrelevant. Indeed, the effectiveness of PRF methods may well depend on the quality of the feedback signal and thus on the effectiveness of the first-stage ranker. This aspect however has received little attention before.

In this paper we control the quality of the feedback signal and measure its impact on a range of PRF methods, including traditional bag-of-words methods (Rocchio), and dense vector-based methods (learnt and not learnt). Our results show the important role the quality of the feedback signal plays on the effectiveness of PRF methods. Importantly, and surprisingly, our analysis reveals that not all PRF methods are the same when dealing with feedback signals of varying quality. These findings are critical to gain a better understanding of the PRF methods and of which and when they should be used, depending on the feedback signal quality, and set the basis for future research in this area. %\todo{Update the findings after finalizing the results section.}

\end{abstract}

%\keywords{Pseudo-Relevance Feedback, Signal Quality, Dense Models, Passage Retrieval}

\maketitle

\section{Introduction and Related Work}

A key assumption in Pseudo-Relevance Feedback is that the top-k documents used as feedback are relevant. Consider for example the scoring formula of the popular Rocchio method~\cite{rocchio1971rocchio}\footnote{The Rocchio method also has allowance for negative feedback -- we removed this here for brevity; we also note that in the PRF setting, negative feedback is often not used.}:

\begin{equation}
	\vec{q'} = \alpha \vec{q} + \beta \frac{1}{|Rel|} \sum_{d_i in Rel} \vec{d_i}  \label{eq:rocchio}
\end{equation}
where, the vectors $\vec{d_i}$ refer to the top-k documents retrieved by the original query $\vec{q}$. Similar treatments are employed by other PRF techniques, both those for bag-of-words models~\cite{azad2019query,robertson2009probabilistic,rocchio1971rocchio,lavrenko2017relevance,lv2009comparative,zhai2001model} and for neural models~\cite{li2021improving,yu2021improving,li2021pseudo}. 

This assumption is often incorrect, i.e. the top-k signal often contain irrelevant documents. Then how do PRF methods behave in the presence of different quality of the relevance signal, e.g. if all $k$ documents are relevant vs. if all of them are not relevant? And do PRF methods differ in their behaviour when examining signals of different quality, e.g., a method that is more effective than another when the relevance signal is of high quality, exhibits large losses compared to the other method when the feedback signal is of poor quality?
These aspects have often been ignored in the PRF literature, and there is no systematic understanding of how PRF behaves depending on the feedback signal quality, nor how results from methods differ depending on the feedback quality. 
However, these are important considerations to make, for a number of reasons. 

First, PRF has been shown to provide mixed effectiveness~\cite{azad2019query}. The factors affecting PRF effectiveness may be many, and certainly include representation choices, PRF depth, and method-specific settings (e.g., for Rocchio, these would be the weights $\alpha$ and $\beta$) ~\cite{tsai2015factors}. In addition, we posit that the feedback signal quality also plays a fundamental role in shaping PRF's effectiveness -- our empirical results reinforce this standing. 
 
Second, PRF is often studied in the context of a standard first-stage retrieval method, commonly BM25, and statements regarding the comparative effectiveness of different PRF methods are made in this context. However, it is unclear whether these statements would be valid if a stronger, or weaker, first-stage retrieval was used. This is important because, in practice, many would transfer the findings from such research into production systems that may differ in terms of the quality of the feedback signal provided to the PRF technique. 

In this paper we provide the first systematic understanding of how feedback signal quality impacts the effectiveness of PRF techniques. We do this in the context of the Rocchio method for bag-of-words models and of two PRF methods for dense retrievers, techniques that have recently gained momentum both in the research literature~\cite{yu2021improving,li2021improving} and in practical adoption~\cite{li2021pseudo}.

\section{Method}

The goal of this study is to evaluate how the quality of feedback signals affects the performance of PRF methods. %-- and to investigate how each retrieval model reacts to different PRF signals.
 To achieve this, we devise two sets of different experiments to control the PRF signal from two different sources. In this section, we outline how we perform PRF with the controlled feedback signals on top of different initial retrieval methods.

\subsection{Controlling the Quality of the Feedback Signal}

In all experiments, we consider three different levels of quality of the feedback signal: strong, moderate, and weak. In the experiments we use the MS MARCO passage ranking dataset~\cite{nguyen2016ms} and the queries from the TREC Deep Learning Track Passage Retrieval Task 2019~\cite{craswell2020overview} (TREC DL 2019) and 2020~\cite{craswell2021overview} (TREC DL 2020); detailed statistics for these datasets are given in Table~\ref{table:stats}. Assessments in these datasets are graded on a 4-point relevance scale -- 0: irrelevant; 1:relevant; 2: highly relevant; 3: perfectly relevant. We define the feedback signal as \textit{strong} if all $k$ passages included in the signal have relevance label 2 or 3. We define the signal as \textit{moderate} if all $k$ passages have label 1; otherwise if all passages have label 0 we define the signal and being \textit{weak}. For simplicity, we do not consider \textit{mixed signals}, where the relevance of the passages in the top $k$ varies, but it is easy to extend this work in that direction. In terms of PRF depth $k$, we study $k=1$ and $k=3$. This choice was made because the ANCE-PRF~\cite{yu2021improving} model checkpoint shared by the original author has been created for $k=3$ and thus not optimised for higher values of $k$. We also note that \citet{yu2021improving} investigated other depths settings from 0 to 5 and found that the checkpoint with $k=3$ provides the highest effectiveness. Furthermore, we highlight that depths values larger than 5-6 are not possible in ANCE-PRF on the MS MARCO corpus because the text of passages beyond those values would be ignored by ANCE-PRF, due to the limited size of input the ANCE encoder accepts.

We first consider the feedback signal obtained from a first-stage retriever. As first stage retrievers, we consider a representative bag-of-words method, BM25~\cite{robertson2009probabilistic}, and three representative dense retrievers methods, namely ANCE~\cite{xiong2020approximate}, TCTv2-HN~\cite{lin2021batch}, and DistillBERT-Balanced~\cite{hofstatter2021efficiently}. Once the initial retrieval is performed (results up to rank position 1,000), we filter the results to remove all unjudged passages. Then, we filter once more to form three distinguished rankings by only considering passages with labels 2 and 3 (for strong signal), 1 (for moderate signal) and 0 (for weak signal). From each set, we then sample 12 passages for each query; if a query does not have 12 passages in one of the three sets (e.g., has less than 12 passages with label 1), then the query is discarded from all sets and ignored for the evaluation. The statistics for the resulting filtered datasets are also reported in Table~\ref{table:stats}. The rationale for choosing 12 passages is as follows. First, we recorded the number of judged passages for each relevant level for each query. From this distribution we then identified the smallest amount of judged passages across any label -- choosing this amount of passages in our experiment would guarantee that every query then has the same amount of unique passages for signal type. However, since the depth $k$ values we experimented with are 1 and 3, we also need to ensure the number of selected passages is a multiple of 1 and 3. This last requirement resulted in identifying 12 as the largest suitable number of passages to select\footnote{Note: we removed 7 out of 36 queries from TREC 2019 (removed query ids: '1124210', '443396', '855410', '1117099', '1037798', '1121709', '131843') and 12 out of 42 queries from TREC 2020 (removed query ids: '1116380', '405163', '42255', '1105792', '1115210', '324585', '1131069', '673670', '336901', '768208', '1030303', '258062'). These queries were removed because for each of these queries at least a label was not sufficiently represented (i.e. less than 10 passages).}.

% \todo{[GZ: Need to say w excluded some queries because had too less labels: which queries?]} \todo{Hang: TREC 2019 has 36 valid queries, 7 queries are omitted, ('1124210', '443396', '855410', '1117099', '1037798', '1121709', '131843'); TREC 2020 has 42 valid queries, 12 queries are omitted, ('1116380', '405163', '42255', '1105792', '1115210', '324585', '1131069', '673670', '336901', '768208', '1030303', '258062')}

When $k=1$, we use the 12 passages for each query to generate 12 runs using for each a different passage from the set as the relevance feedback signal. Then, the runs for a query are averaged and results are reported. Thus, for each query, we have 3 main results, one for each level of the feedback signal (i.e. strong, moderate, weak) -- each of these was obtained by averaging the results obtained from 12 instances of the corresponding signal.

The process when $k=3$ is similar, apart that, for each query, we randomly split the 12 passages into 4 groups, each containing 3 feedback passages. For each query, we take all 4 groups, and perform PRF, to produce 4 runs for a single query, then we average the performance of these 4 runs to get the final performance of for that query, on a specific level of feedback quality. Thus, for each query, we have 3 main results, one for each level of the feedback signal (i.e. strong, moderate, weak) -- each of these was obtained by averaging the results obtained from 4 instances of the corresponding signal and each of these contained 3 passages.

We then repeat the settings above, but sampling passages from the qrels~\footnote{The file containing the relevance assessments.} rather than from the baseline runs. We do this to remove any influence of a strong or weak first-stage retrieval on our findings. %We do this because \todo{xxx Hang, do you have notes of the reasons?}. 
Queries that were excluded before because the rankings contained less than 12 passages of any given label are also ignored here.

\begin{table}[]
\centering
\caption{Statistics of the two datasets considered in our experiments. The statistics of the datasets after we remove the queries that do not have enough judged passages are labeled with (Filtered). We use the Filtered datasets in our experiments.}
\resizebox{\columnwidth}{!}{%
\begin{tabular}{lrrrrr}
\toprule
                        & \#Queries & \#Passages        & Avg \#J/Q & \#Judgements \\ \midrule
TREC DL 2019            & 43  & 8,841,823  & 215.3    & 9,260 \\
TREC DL 2019 (Filtered) & 36  & 8,841,823  & 217.7    & 7,838  \\ \midrule
TREC DL 2020            & 54  & 8,841,823  & 210.9    & 11,386 \\
TREC DL 2020 (Filtered) & 42  & 8,841,823  & 212.8    & 8,936  \\\bottomrule
\end{tabular}
}
\label{table:stats}
\end{table}

\subsection{Considered PRF Methods}

While there are many methods of retrieval and PRF being proposed in the literature, in this first investigation we consider a subset of these methods that allows us to understand what the impact of feedback quality is on PRF effectiveness with respect to representation type, i.e. bag-of-words vs. dense vectors, and PRF type, i.e. learnt vs. not learnt. 

Based on this, we decided to use BM25 as a representative bag-of-words method, noting that differences with other methods such as Language Modelling are often not substantial, along with ANCE~\cite{xiong2020approximate}, TCTv2-HN~\cite{lin2021batch}, and DistillBERT-Balanced~\cite{hofstatter2021efficiently} as representative dense retrievers. Note that ANCE is based on RoBERTA, TCTv2-HN on BERT and DistillBERT-Balanced on a reduced version of BERT (learnt with knowledge distillation), and thus do differ to some extent in terms of representation. 

Similarly, we selected the ANCE-PRF method~\cite{yu2021improving} and its extensions to TCTv2-HN and DistillBERT-Balanced by \citet{li2021improving}, as representative \textit{learnt} PRF methods. In these methods, in fact, a PRF encoder is fine-tuned to the relevance feedback task. We note that bag-of-words models do not have a corresponding complex trainable method (often tuning is performed but involves optimizing one or a handful of parameters, not the millions of parameters in the considered transformer-based models). We then selected the Vector-PRF method by \citet{li2021pseudo}; specifically we used the Rocchio variant of their method, which follows the general Rocchio PRF formula of Equation~\ref{eq:rocchio}, but where the vectors are the actual dense representations from the dense encoders used for the first stage retrieval. The parameters in these methods are only two ($\alpha, \beta$) and we set them to the values used in previous work~\cite{li2021pseudo}. The method can be applied on top of any dense retriever, and we apply it to the 3 dense retrievers considered here.
 This method has an obvious correspondence in the bag-of-words space: it's the original Rocchio method -- thus we consider Rocchio PRF on top of BM25, rather than the more popular RM3 method, to have a direct comparison between bag-of-words and dense retrievers under the same PRF strategy.

For all methods, be it bag-of-words or dense retrievers, learnt PRF (a.k.a. ANCE-PRF and derivatives) or Vector-PRF, we use the implementations available in  Anserini/Pyserini~\cite{yang2018anserini,lin2021pyserini} and the checkpoints made available by the corresponding authors of the techniques. We implement Rocchio PRF on top of the bag-of-words model in Pyserini and add this implementation to the GitHub repository associated with our paper\footnote{\url{https://github.com/ielab/Noise-PRF}}.

\section{Results}

%All the results in this section are presented on the TREC DL 2019 and TREC DL 2020 filtered query set (36 queries in TREC DL 2019, 42 queries in TREC DL 2020).

\begin{table*}[t]
	\centering
	%\footnotesize
	\caption{Effectiveness of PRF methods across different representations and PRF signal qualities. $R$ stands for the Rocchio PRF method for bag-of-words, baselines are the PRF runs without control of the PRF signal quality (i.e., standard PRF on top $k$ retrieved documents). For each signal quality, the PRF models are divided into three categories: Rocchio PRF on bag-of-words, VectorPRF-Rocchio on dense retrievers, and trained PRF on dense retrievers. Statistical significance analysis is performed using two-tailed paired Student's ttest with Bonferroni correction; significant differences are marked with $\dagger$.}
	\resizebox{!}{.2\paperheight}{%
		\begin{tabular}{cl|rrrrrrr|rrrrrrr}
			&
			\multirow{2}{*}{Models} &
			\multicolumn{7}{c}{TREC DL 2019} &
			\multicolumn{7}{c}{TREC DL 2020} \\ \cmidrule{3-16}
			\multicolumn{1}{l}{} &
			&
			\multicolumn{1}{c}{MAP} &
			\multicolumn{1}{l}{RR} &
			\multicolumn{1}{l}{R@1000} &
			\multicolumn{1}{l}{nDCG@1} &
			\multicolumn{1}{l}{nDCG@3} &
			\multicolumn{1}{l}{nDCG@10} &
			\multicolumn{1}{l}{nDCG@100} &
			\multicolumn{1}{l}{MAP} &
			\multicolumn{1}{l}{RR} &
			\multicolumn{1}{l}{R@1000} &
			\multicolumn{1}{l}{nDCG@1} &
			\multicolumn{1}{l}{nDCG@3} &
			\multicolumn{1}{l}{nDCG@10} &
			\multicolumn{1}{l}{nDCG@100} \\ \midrule
			\multirow{7}{*}{\rotatebox[origin=c]{90}{\textsc{Baseline}}} &
			BM25 &
			0.2697 &
			0.7044 &
			0.7687 &
			0.5972 &
			0.5298 &
			0.4971 &
			0.4945 &
			0.2870  &
			0.6531 &
			0.7938 &
			0.5595 &
			0.5155 &
			0.4959 &
			0.4959 \\
			&
			ANCE &
			0.3908 &
			0.8501 &
			0.8031 &
			0.7222 &
			0.7022 &
			0.6767 &
			0.5860  &
			0.4047 &
			0.8275 &
			0.7804 &
			0.7619 &
			0.7479 &
			0.6806 &
			0.5670\\
			&
			TCTV2-HN+ &
			0.4676 &
			0.8788 &
			0.8794 &
			0.8009 &
			0.7488 &
			0.7309 &
			0.6631 &
			0.4047 &
			0.8275 &
			0.7804 &
			0.7619 &
			0.7479 &
			0.6806 &
			0.5670\\
			&
			DistilBERT &
			0.4832 &
			0.8763 &
			0.8905 &
			0.7454 &
			0.7466 &
			0.7319 &
			0.6698 &
			0.4742 &
			0.8677 &
			0.8770 &
			0.7778 &
			0.7887 &
			0.7207 &
			0.6382 \\
			&
			BM25+R &
			0.2350$^{\dagger}$ &
			0.6328 &
			0.8044 &
			0.5000$^{\dagger}$ &
			0.4963 &
			0.4483 &
			0.4318$^{\dagger}$ &
			0.1936$^{\dagger}$ &
			0.5198$^{\dagger}$ &
			0.7447 &
			0.4206$^{\dagger}$ &
			0.4246$^{\dagger}$ &
			0.3864$^{\dagger}$ &
			0.3734$^{\dagger}$ \\ \cmidrule{2-16}
			&
			ANCE+VPRF-R &
			0.4300$^{\dagger}$ &
			0.8177 &
			0.8179 &
			0.7037 &
			0.7023 &
			0.6790 &
			0.6202$^{\dagger}$ &
			0.4220$^{\dagger}$ &
			0.8377 &
			0.7958 &
			0.7540 &
			0.7472 &
			0.6841 &
			0.5760 \\
			&
			TCTV2+VPRF-R &
			0.4949$^{\dagger}$ &
			0.8682 &
			0.8942 &
			0.7917 &
			0.7464 &
			0.7406 &
			0.6876 &
			0.4904$^{\dagger}$ &
			0.8321 &
			0.8655$^{\dagger}$ &
			0.7817 &
			0.7592 &
			0.7144 &
			0.6274$^{\dagger}$ \\
			&
			DistilBERT+VPRF-R &
			0.5156$^{\dagger}$ &
			0.8606 &
			0.8928 &
			0.7731 &
			0.7411 &
			0.7387 &
			0.6897 &
			0.4974$^{\dagger}$ &
			0.8899 &
			0.9101$^{\dagger}$ &
			0.8135 &
			0.7973 &
			0.7513 &
			0.6535 \\ \cmidrule{2-16}
			&
			ANCE-PRF &
			0.4423$^{\dagger}$ &
			0.8721 &
			0.8293 &
			0.7361 &
			0.7204 &
			0.7074 &
			0.6270$^{\dagger}$ &
			0.4340$^{\dagger}$ &
			0.8881$^{\dagger}$ &
			0.8286 &
			0.8571$^{\dagger}$ &
			0.7792 &
			0.7275 &
			0.5897 \\
			&
			TCTV2-PRF &
			0.4901$^{\dagger}$ &
			0.8615 &
			0.8888 &
			0.7500 &
			0.7606 &
			0.7456 &
			0.6802 &
			0.4864$^{\dagger}$ &
			0.8774 &
			0.8562$^{\dagger}$ &
			0.8254$^{\dagger}$ &
			0.8038 &
			0.7331 &
			0.6252$^{\dagger}$ \\
			&
			DistilBERT-PRF &
			0.4996 &
			0.8588 &
			0.8968 &
			0.7546 &
			0.7648 &
			0.7386 &
			0.6778 &
			0.4860 &
			0.8810 &
			0.8777 &
			0.7976 &
			0.7803 &
			0.7306 &
			0.6293 \\ \midrule \midrule
			\multirow{7}{*}{\rotatebox[origin=c]{90}{\textsc{Strong Signal}}} &
			BM25+R &
			0.3706$^{\dagger}$ &
			0.8334$^{\dagger}$ &
			0.8412 &
			0.6505 &
			0.6536$^{\dagger}$ &
			0.6076$^{\dagger}$ &
			0.5578$^{\dagger}$ &
			0.3936$^{\dagger}$ &
			0.8859$^{\dagger}$ &
			0.8422 &
			0.7695$^{\dagger}$ &
			0.7040$^{\dagger}$ &
			0.6411$^{\dagger}$ &
			0.5566 \\ \cmidrule{2-16}
			&
			ANCE+VPRF-R &
			0.5119$^{\dagger}$ &
			0.8816$^{\dagger}$ &
			0.8531$^{\dagger}$ &
			0.7708 &
			0.7690 &
			0.7511 &
			0.6749$^{\dagger}$ &
			0.5259$^{\dagger}$ &
			0.9164$^{\dagger}$ &
			0.8377 &
			0.8132 &
			0.8158 &
			0.7670$^{\dagger}$ &
			0.6398$^{\dagger}$ \\
			&
			TCTV2+VPRF-R &
			0.5769$^{\dagger}$ &
			0.9136 &
			0.9292 &
			0.8002 &
			0.7957 &
			0.7773 &
			0.7366$^{\dagger}$ &
			0.6018$^{\dagger}$ &
			0.9484$^{\dagger}$ &
			0.9142$^{\dagger}$ &
			0.8323 &
			0.8227 &
			0.7925$^{\dagger}$ &
			0.6918$^{\dagger}$ \\
			&
			DistilBERT+VPRF-R &
			0.5931$^{\dagger}$ &
			0.9410 &
			0.9336 &
			0.7967 &
			0.8068 &
			0.7971 &
			0.7433 &
			0.6022$^{\dagger}$ &
			0.9738$^{\dagger}$ &
			0.9255 &
			0.8462 &
			0.8260 &
			0.7955$^{\dagger}$ &
			0.7030 \\ \cmidrule{2-16}
			&
			ANCE-PRF &
			0.4907$^{\dagger}$ &
			0.9060 &
			0.8388 &
			0.7986 &
			0.7722 &
			0.7484 &
			0.6583$^{\dagger}$ &
			0.4798$^{\dagger}$ &
			0.8771 &
			0.8241 &
			0.7798 &
			0.7685 &
			0.7249 &
			0.6078 \\
			&
			TCTV2-PRF &
			0.5326$^{\dagger}$ &
			0.8807 &
			0.9134 &
			0.7654 &
			0.7641 &
			0.7600 &
			0.7102 &
			0.5348$^{\dagger}$ &
			0.9220 &
			0.8780$^{\dagger}$ &
			0.8307 &
			0.8223 &
			0.7683$^{\dagger}$ &
			0.6512$^{\dagger}$ \\
			&
			DistilBERT-PRF &
			0.5408$^{\dagger}$ &
			0.8954 &
			0.9124 &
			0.7963 &
			0.7821 &
			0.7630 &
			0.7041 &
			0.5264$^{\dagger}$ &
			0.9082 &
			0.8915 &
			0.8185 &
			0.8084 &
			0.7542 &
			0.6561 \\ \midrule \midrule
			\multirow{7}{*}{\rotatebox[origin=c]{90}{\textsc{Moderate Signal}}} &
			BM25+R &
			0.2641 &
			0.5738$^{\dagger}$ &
			0.7979 &
			0.4842$^{\dagger}$ &
			0.4899 &
			0.4871 &
			0.4975 &
			0.1960$^{\dagger}$ &
			0.3855$^{\dagger}$ &
			0.7956 &
			0.3740$^{\dagger}$ &
			0.3821$^{\dagger}$ &
			0.4039$^{\dagger}$ &
			0.4358$^{\dagger}$ \\ \cmidrule{2-16}
			&
			ANCE+VPRF-R &
			0.4365 &
			0.8278 &
			0.8477 &
			0.7161 &
			0.6951 &
			0.6845 &
			0.6444 &
			0.3541$^{\dagger}$ &
			0.6839$^{\dagger}$ &
			0.8174 &
			0.6177$^{\dagger}$ &
			0.6286$^{\dagger}$ &
			0.6091 &
			0.5738 \\
			&
			TCTV2+VPRF-R &
			0.4675 &
			0.7999 &
			0.9088 &
			0.7164$^{\dagger}$ &
			0.7060 &
			0.7052 &
			0.6861 &
			0.3754 &
			0.5763$^{\dagger}$ &
			0.8830$^{\dagger}$ &
			0.5321$^{\dagger}$ &
			0.5665$^{\dagger}$ &
			0.5792$^{\dagger}$ &
			0.6124 \\
			&
			DistilBERT+VPRF-R &
			0.4847 &
			0.8143 &
			0.9193 &
			0.7029 &
			0.7049 &
			0.7075 &
			0.6981 &
			0.4075$^{\dagger}$ &
			0.6412$^{\dagger}$ &
			0.8889 &
			0.5747$^{\dagger}$ &
			0.6140$^{\dagger}$ &
			0.6440 &
			0.6315 \\ \cmidrule{2-16}
			&
			ANCE-PRF &
			0.4369$^{\dagger}$ &
			0.7740 &
			0.8324 &
			0.6620 &
			0.6905 &
			0.6936 &
			0.6385 &
			0.3703 &
			0.6882$^{\dagger}$ &
			0.8143 &
			0.6359$^{\dagger}$ &
			0.6409$^{\dagger}$ &
			0.6302 &
			0.5627 \\
			&
			TCTV2-PRF &
			0.4841 &
			0.8550 &
			0.8977 &
			0.7558 &
			0.7403 &
			0.7338 &
			0.6915 &
			0.4684$^{\dagger}$ &
			0.8540 &
			0.8583 &
			0.7791 &
			0.7545 &
			0.7113 &
			0.6355$^{\dagger}$ \\
			&
			DistilBERT-PRF &
			0.5049 &
			0.8779 &
			0.9069 &
			0.7755 &
			0.7472 &
			0.7369 &
			0.6900 &
			0.4701 &
			0.8215 &
			0.8816 &
			0.7229 &
			0.7499 &
			0.7018 &
			0.6340 \\ \midrule \midrule
			\multirow{7}{*}{\rotatebox[origin=c]{90}{\textsc{Weak Signal}}} &
			BM25+R &
			0.1957$^{\dagger}$ &
			0.3917$^{\dagger}$ &
			0.7641 &
			0.2118$^{\dagger}$ &
			0.2518$^{\dagger}$ &
			0.2902$^{\dagger}$ &
			0.3643$^{\dagger}$ &
			0.1839$^{\dagger}$ &
			0.3712$^{\dagger}$ &
			0.7433 &
			0.2153$^{\dagger}$ &
			0.2389$^{\dagger}$ &
			0.2866$^{\dagger}$ &
			0.3506$^{\dagger}$ \\ \cmidrule{2-16}
			&
			ANCE+VPRF-R &
			0.3915 &
			0.7886 &
			0.8130 &
			0.6713 &
			0.6526 &
			0.6480 &
			0.5836 &
			0.3180$^{\dagger}$ &
			0.6949$^{\dagger}$ &
			0.7832 &
			0.5575$^{\dagger}$ &
			0.5609$^{\dagger}$ &
			0.5250$^{\dagger}$ &
			0.4968$^{\dagger}$ \\
			&
			TCTV2+VPRF-R &
			0.4408 &
			0.8036 &
			0.8875 &
			0.6921$^{\dagger}$ &
			0.6608 &
			0.6692 &
			0.6351 &
			0.3440$^{\dagger}$ &
			0.5926$^{\dagger}$ &
			0.8352 &
			0.4484$^{\dagger}$ &
			0.4785$^{\dagger}$ &
			0.4792$^{\dagger}$ &
			0.5031 \\
			&
			DistilBERT+VPRF-R &
			0.4441 &
			0.8022 &
			0.8970 &
			0.6667$^{\dagger}$ &
			0.6622$^{\dagger}$ &
			0.6561 &
			0.6371 &
			0.3904$^{\dagger}$ &
			0.7441$^{\dagger}$ &
			0.8603 &
			0.5903$^{\dagger}$ &
			0.5873$^{\dagger}$ &
			0.5702$^{\dagger}$ &
			0.5555$^{\dagger}$ \\ \cmidrule{2-16}
			&
			ANCE-PRF &
			0.3929 &
			0.7121$^{\dagger}$ &
			0.8159 &
			0.5232$^{\dagger}$ &
			0.5733$^{\dagger}$ &
			0.5906$^{\dagger}$ &
			0.5791 &
			0.3594 &
			0.7185$^{\dagger}$ &
			0.7911 &
			0.6181$^{\dagger}$ &
			0.6142$^{\dagger}$ &
			0.5935 &
			0.5280 \\
			&
			TCTV2-PRF &
			0.4705 &
			0.8487 &
			0.8890 &
			0.7315 &
			0.7110 &
			0.7053 &
			0.6582 &
			0.4703$^{\dagger}$ &
			0.8827 &
			0.8551$^{\dagger}$ &
			0.7748 &
			0.7488 &
			0.7001 &
			0.6079 \\
			&
			DistilBERT-PRF &
			0.5047 &
			0.8757 &
			0.9002 &
			0.7639 &
			0.7386 &
			0.7262 &
			0.6789 &
			0.4746 &
			0.8626 &
			0.8747 &
			0.7474 &
			0.7675 &
			0.6991 &
			0.6213 \\ \bottomrule
	\end{tabular}}%}
	\label{tab:repre}
\end{table*}

\subsection{Signal Quality and PRF Methods}

	%PRF Signal Quality Impact on Effectiveness}

%Firstly, we investigate how the PRF signal quality impact the effectiveness on all of the PRF models in our experiments. Figure~\ref{figure} reports MAP, Reciprocal Rank (RR), nDCG@{1,3,10,100}, Recall@1000 (R@1000) for the effectiveness of each model with different PRF signal qualities.

First, we investigate the interplay between signal quality and the different PRF methods. Table~\ref{tab:repre} reports MAP, Reciprocal Rank (RR), nDCG@{1,3,10,100}, Recall@1000 (R@1000) for the effectiveness of each model with different PRF signal qualities. For simplicity, we only show model effectiveness with PRF depth 3, since either PRF depth 1 or 3 show similar trends.

%\paragraph{Traditional PRF Method} We use Rocchio~\cite{rocchio1971rocchio} on top of traditional bag-of-words retrieval model BM25. Uncontrolled PRF signals only improve R@1000 with PRF depth 1,3 and nDCG@3 with PRF depth 1 on TREC DL 2019; only nDCG@1 is on par with the BM25 baseline on TREC DL 2020; all other evaluation metrics dropped after applying PRF with depth 1 or 3. However, when we control the quality of the PRF signals, strong signals (judged passages with relevance label 2 and 3), all evaluation metrics on both datasets are substantially increased compared to the baseline BM25 with either PRF depth 1 or 3. When we set the PRF signal quality to be moderate (judged passages with relevance label 1), only marginal improvements can be observed over R@1000, with PRF depth 1, 3, and nDCG@100 with PRF depth 3 on TREC DL 2019; on TREC DL 2020, only marginal improvements can be observed over R@1000, all other evaluation metrics on both datasets are worse than the baseline BM25+Rocchio without controlling the PRF signals. If we only use the weak PRF signals (judged passages with relevance label 0), all metrics are much worse than the baseline model without controlling the signal quality or with moderate signals or with strong signals, some losses are even larger than 60\% compared to the baseline model.

\paragraph{Rocchio PRF}
We use Rocchio~\cite{rocchio1971rocchio} on top of the bag-of-words retrieval model BM25 as well as an adaptation of the Rocchio method for dense retrievers, called Vector PRF~\cite{li2021pseudo}. For the dense retrievers, we applied Vector PRF on top of ANCE, TCTV2-HN and DistillBERT. The parameter settings are presented in Table~\ref{tab:para}. For BM25, the Rocchio parameters were set to $\alpha = 0.75$ and $\beta=0.15$, following previous literature.
For all dense methods, they were set to $\alpha = 0.6$ and $\beta=0.4$ on TREC DL 2019 and on 2020 (only when $k=3$) and to $\alpha = 0.5$ and $\beta=0.5$ on TREC DL 2020 when $k=1$. These choices were made based on the results from \citet{li2021pseudo}.

\begin{table}[]
\caption{The Rocchio parameter settings for both datasets, with different PRF depths and different models.}
\begin{tabularx}{\columnwidth}{c|c|c|cc}
                                  & Depth                &                       & TREC DL 2019 & TREC DL 2020 \\ \toprule
\multirow{2}{*}{BOW}              & \multirow{2}{*}{all} & $\alpha$ & 0.75         & 0.75         \\ 
                                  &                      & $\beta$  & 0.15         & 0.15         \\ \midrule
\multirow{4}{*}{Dense Retrievers} & \multirow{2}{*}{1}   & $\alpha$ & 0.6          & 0.5          \\
                                  &                      & $\beta$  & 0.4          & 0.5          \\ \cmidrule{2-5}
                                  & \multirow{2}{*}{3}   & $\alpha$ & 0.6          & 0.6          \\
                                  &                      & $\beta$  & 0.4          & 0.4        \\ \bottomrule 
\end{tabularx}
\label{tab:para}
\end{table}

Although somewhat tuned, then, this Rocchio PRF method was not ``learnt'' (as opposed to the learnt PRF methods below).

When bag-of-words are used, the PRF signal extracted from the first stage without further filtering (uncontrolled PRF signal) only improves R@1000 with PRF depth 1,3 and nDCG@3 with PRF depth 1 on TREC DL 2019; nDCG@1 is on par with the BM25 baseline on TREC DL 2020; and all other metrics exhibit drops after the use of PRF.
However, when we control the quality of the PRF signals, strong signals substantially enhance the effectiveness over all metrics and datasets; moderate signals marginally improve R@1000; weak signals hurt the effectiveness significantly across all metrics, with some losses even larger than 60\% compared to the BM25 baseline model.

When dense retrievers are used, the uncontrolled PRF signal gives rise to improvements across the majority of metrics on both datasets.
With strong PRF signals, the improvements are significant across all metrics on both datasets, except for TCTV2+VPRF-Rocchio in nDCG@1 on TREC DL 2019 with PRF depth 3.
When the PRF signals are moderate, ANCE+VPRF-Rocchio still achieves significant improvements in terms of MAP, R@1000, and nDCG@100; for TCTV2+VPRF-Rocchio, however, effectiveness decreases quickly compared to the strong signals, resulting in most metrics being now significantly worse than the baseline models on both datasets. A similar behaviour occurs for DistilBERT+VPRF-Rocchio.
When the weak PRF signals are used, improvements in TREC DL 2019 are observed only for R@1000 with PRF depth 3 for all three models; improvements in TREC DL 2020 are observed only for ANCE+VPRF-Rocchio with PRF depth 3 and TCTV2+VPRF-Rocchio with PRF depths 1 and 3. With DistilBERT+VPRF-Rocchio in TREC DL 2020 all metrics are worse than the baseline and some losses are larger than 40\%.

\paragraph{Learnt PRF}
We use ANCE-PRF~\cite{yu2021improving} and its variants TCTV2-PRF and DistilBERT-PRF~\cite{li2021improving} as example of learnt PRF methods on dense representations. Bag-of-words representations do not have an equivalent, heavily learnt PRF method.

When not controlling the quality of PRF signals, all three models substantially improve the respective models without PRF on most metrics across both datasets.
When the PRF signals is strong, all three models improve significantly more over the baseline models on both datasets, except TCTV2-PRF for nDCG@1 on TREC DL 2019.
By using moderate signals, for all three models larger improvements only occur for deep metrics, such as MAP, nDCG@100, and R@1000. For other metrics instead, effectiveness is either on par or worse than the baseline models (without PRF), on both datasets.
For weak signals, marginal improvements can still be observed for deep metrics, but these are much smaller than for other PRF signal qualities, while losses are abundant and some are larger than 20\%.

%For trained dense PRF method, we use ANCE-PRF~\cite{yu2021improving}, TCTV2-PRF~\cite{li2021improving}, DistilBERT-PRF~\cite{li2021improving}. Because in these papers, only models with PRF depth 3 are provided, so for this approach, we only consider PRF depth 3. When we are not controlling the quality of PRF signals, all three models substantially improve the baseline models on most metrics across both datasets. When we set the PRF signals to strong, all three models improve more over the baseline models on both datasets, except the nDCG@1 on TREC DL 2019 with TCTV2-PRF model. By using the moderate signals, larger improvements only occur over deep metrics, such as MAP, nDCG@100, and R@1000 for all three models, for other metrics the performance is either on par or worse than the baseline models on both datasets. For weak signals, marginal improvements can still be observed over deep metrics, but much smaller than other PRF signal qualities, some losses are larger than 20\%.

In conclusion, with strong PRF signals, Rocchio PRF approaches, either BM25+Rocchio or Vector PRF, can improve the performance across all metrics on both datasets. However, when we change to use only moderate signals, BM25+Rocchio only can marginally improve deep recall, where ANCE+VPRF-Rocchio is more resilient to this change and show substantial improvements over all deep metrics, TCTV2+VPRF-Rocchio and DistilBERT+VPRF-Rocchio, on the other hand, also drops quickly. For using weak signals, BM25+Rocchio suffers more than 60\% loss on several metrics compares to BM25, marginal improvements can be observed only for ANCE+VPRF-Rocchio and TCTV2+VPRF-Rocchio, where DistilBERT+VPRF-Rocchio still suffers from substantial loss. For the learnt PRF approaches, all three models show a more stable resilient of signal quality change, even with weak signal, the worst performance are just about 20\% lower than the baseline.

%\vspace{-10pt}
\subsection{Signal Quality and Representations}

% \usepackage{multirow}

Next, we investigate how representations from different models impact effectiveness. For this analysis, we only consider the Rocchio PRF method (called Vector-PRF or VPRF-Rocchio for dense retrievers), as this is the only PRF method for which we have both bag-of-words and dense representations. We again refer to the results in Table~\ref{tab:repre}.

For the bag-of-words representation (BM25 + Rocchio), effectiveness drops very quickly when moving from a strong signal to a weak signal: losses at times reach 80\% for some metrics. This trend is observed across all datasets and metrics.

We now consider dense representations.
ANCE+VPRF-Rocchio exhibits more stable behaviour with respect to changes of feedback signal quality than when bag-of-words are used: losses in the worst conditions are up to only 30\%. However, TCTV2+VPRF-Rocchio and DistilBERT+VPRF-Rocchio show instead quite unstable patterns when changing the PRF signal quality: the methods suffer losses of more than 50\%  on some metrics when changing the feedback signal from strong  to weak. 

%For trained dense PRF methods, ANCE-PRF is similar to ANCE+VPRF-Rocchio but with slightly larger losses, as well as TCTV2-PRF and DistilBERT-PRF, all three trained PRF models are showing similar stability with different PRF signal qualities.

Our results suggest that better underlying representations, i.e. dense representations in place of bag-of-words representations, lead the same PRF technique to higher effectiveness, and this is regardless of the feedback signal quality. In fact, even with feedback signal of weak quality, losses obtained by the PRF mechanism on dense representations are lower than those obtained on bag-of-words representations. Differences do still exist however across the different dense representations, at least in the extent of th relative gains and losses depending on the quality of the PRF signal.

%From our results, we can see that with better vector representations, the PRF mechanism works better. The traditional PRF method, Rocchio, generates the vector representations for query and passages without training, hence it does not have a good vector representations for the PRF mechanism, which makes the effectiveness unstable with different PRF signal qualities. Similar for ANCE+VPRF-Rocchio, TCTV2+VPRF-Rocchio, and DistilBERT+VPRF-Rocchio, these models have a much smaller loss compared to the traditional PRF method, because the vector representations of queries and passages are from the base models (ANCE, TCTV2, DistilBERT), which are trained to contain more information in the vectors. However, these PRF methods heavily depend on the base models' vector representations, so if the base model's vector representation is not good enough, then the effectiveness after PRF is also unstable, such as TCTV2+VPRF-Rocchio and DistilBERT+VPRF-Rocchio. On the other hand, with the trained PRF methods (ANCE-PRF, TCTV2-PRF, and DistilBERT-PRF), they all suffer minimal losses by changing the PRF signal quality from strong to weak, which suggests that with training, it can improve the vector representations of the base models, and keep the effectiveness more stable with different PRF signal qualities.

\section{Conclusion}

In this paper we conducted a systematic investigation of how the feedback signal quality impacts the effectiveness of pseudo relevance feedback for passage retrieval. We demonstrated that the strength of the PRF signals has a high impact on effectiveness; strong signals achieve higher gains in effectiveness, while weak signals hurt the effectiveness. However, we showed that the stability in performance differs from one PRF method to another. For instance, \textit{learnt} PRF methods are more resilient to weak signals (noise) than not-learnt methods (e.g. Rocchio on either bag-of-words representations or dense retrievers -- called VectorPRF Rocchio). We also showed that, under the same PRF method, dense representations are better than bag-of-words representations across all spectrum of feedback signal quality.

Our investigation is not without limitations. Importantly, we did not consider mixed signals (i.e., where the relevance of passages in the top $k$ varies) and a broader set of PRF depths $k$. Mixed signals were not considered in this initial work so as to have a clear control of the signals and facilitate our investigation and results interpretations. However, in future work we plan to extend our analysis to more complex signals, including larger samples. In terms of feedback depth, we only studied $k=1$ and $k=3$ -- although these are often popular settings, especially for the passage retrieval task, certainly they are not the only possible\footnote{While some dense retrieval based PRF methods are limited in terms of the maximum number of passage they can consider as feedback~\cite{yu2021improving,li2021improving}, others are not~\cite{li2021pseudo}.}. In addition, more PRF methods could have been investigated, including other neural PRF methods, e.g., ColBERT-PRF~\cite{wang2021pseudo}.

\section*{Acknowledgements} This research is funded by the Grain Research and Development Corporation (GRDC), project AgAsk (UOQ2003-009RTX).

%We believe that the systematic investigation we provide in this paper regarding how the quality of the feedback signal influences the effectiveness of PRF methods can motivate others to consider this aspect when investigating a new PRF technique.  

\vfill\eject

\bibliographystyle{ACM-Reference-Format}
\interlinepenalty=10000
\bibliography{impact-noise-PRF}


\begin{thebibliography}{19}

%%% the Web address.

\ifx \showCODEN    \undefined \def \showCODEN     #1{\unskip}     \fi
\ifx \showDOI      \undefined \def \showDOI       #1{#1}\fi
\ifx \showISBNx    \undefined \def \showISBNx     #1{\unskip}     \fi
\ifx \showISBNxiii \undefined \def \showISBNxiii  #1{\unskip}     \fi
\ifx \showISSN     \undefined \def \showISSN      #1{\unskip}     \fi
\ifx \showLCCN     \undefined \def \showLCCN      #1{\unskip}     \fi
\ifx \shownote     \undefined \def \shownote      #1{#1}          \fi
\ifx \showarticletitle \undefined \def \showarticletitle #1{#1}   \fi
\ifx \showURL      \undefined \def \showURL       {\relax}        \fi
% invisible to TeX
\providecommand\bibfield[2]{#2}
\providecommand\bibinfo[2]{#2}
\providecommand\natexlab[1]{#1}
\providecommand\showeprint[2][]{arXiv:#2}

\bibitem[\protect\citeauthoryear{Azad and Deepak}{Azad and Deepak}{2019}]%
        {azad2019query}
\bibfield{author}{\bibinfo{person}{Hiteshwar~Kumar Azad} {and}
  \bibinfo{person}{Akshay Deepak}.} \bibinfo{year}{2019}\natexlab{}.
\newblock \showarticletitle{Query expansion techniques for information
  retrieval: a survey}.
\newblock \bibinfo{journal}{\emph{Information Processing \& Management}}
  \bibinfo{volume}{56}, \bibinfo{number}{5} (\bibinfo{year}{2019}),
  \bibinfo{pages}{1698--1735}.
\newblock


\bibitem[\protect\citeauthoryear{Craswell, Mitra, Yilmaz, Campos, and
  Voorhees}{Craswell et~al\mbox{.}}{2020}]%
        {craswell2020overview}
\bibfield{author}{\bibinfo{person}{Nick Craswell}, \bibinfo{person}{Bhaskar
  Mitra}, \bibinfo{person}{Emine Yilmaz}, \bibinfo{person}{Daniel Campos},
  {and} \bibinfo{person}{Ellen~M Voorhees}.} \bibinfo{year}{2020}\natexlab{}.
\newblock \showarticletitle{Overview of the TREC 2019 Deep Learning Track}. In
  \bibinfo{booktitle}{\emph{Text REtrieval Conference, TREC}}.
\newblock


\bibitem[\protect\citeauthoryear{Craswell, Mitra, Yilmaz, Campos, and
  Voorhees}{Craswell et~al\mbox{.}}{2021}]%
        {craswell2021overview}
\bibfield{author}{\bibinfo{person}{Nick Craswell}, \bibinfo{person}{Bhaskar
  Mitra}, \bibinfo{person}{Emine Yilmaz}, \bibinfo{person}{Daniel Campos},
  {and} \bibinfo{person}{Ellen~M Voorhees}.} \bibinfo{year}{2021}\natexlab{}.
\newblock \showarticletitle{Overview of the TREC 2020 Deep Learning Track}. In
  \bibinfo{booktitle}{\emph{Text REtrieval Conference, TREC}}.
\newblock


\bibitem[\protect\citeauthoryear{Hofst{\"a}tter, Lin, Yang, Lin, and
  Hanbury}{Hofst{\"a}tter et~al\mbox{.}}{2021}]%
        {hofstatter2021efficiently}
\bibfield{author}{\bibinfo{person}{Sebastian Hofst{\"a}tter},
  \bibinfo{person}{Sheng-Chieh Lin}, \bibinfo{person}{Jheng-Hong Yang},
  \bibinfo{person}{Jimmy Lin}, {and} \bibinfo{person}{Allan Hanbury}.}
  \bibinfo{year}{2021}\natexlab{}.
\newblock \showarticletitle{Efficiently Teaching an Effective Dense Retriever
  with Balanced Topic Aware Sampling}.
\newblock \bibinfo{journal}{\emph{arXiv preprint arXiv:2104.06967}}
  (\bibinfo{year}{2021}).
\newblock


\bibitem[\protect\citeauthoryear{Lavrenko and Croft}{Lavrenko and
  Croft}{2017}]%
        {lavrenko2017relevance}
\bibfield{author}{\bibinfo{person}{Victor Lavrenko} {and}
  \bibinfo{person}{W~Bruce Croft}.} \bibinfo{year}{2017}\natexlab{}.
\newblock \showarticletitle{Relevance-Based Language Models}. In
  \bibinfo{booktitle}{\emph{ACM SIGIR Forum}}, Vol.~\bibinfo{volume}{51}. ACM
  New York, NY, USA, \bibinfo{pages}{260--267}.
\newblock


\bibitem[\protect\citeauthoryear{Li, Mourad, Zhuang, Koopman, and Zuccon}{Li
  et~al\mbox{.}}{2021}]%
        {li2021pseudo}
\bibfield{author}{\bibinfo{person}{Hang Li}, \bibinfo{person}{Ahmed Mourad},
  \bibinfo{person}{Shengyao Zhuang}, \bibinfo{person}{Bevan Koopman}, {and}
  \bibinfo{person}{Guido Zuccon}.} \bibinfo{year}{2021}\natexlab{}.
\newblock \showarticletitle{Pseudo relevance feedback with deep language models
  and dense retrievers: Successes and pitfalls}.
\newblock \bibinfo{journal}{\emph{arXiv preprint arXiv:2108.11044}}
  (\bibinfo{year}{2021}).
\newblock


\bibitem[\protect\citeauthoryear{Li, Zhuang, Mourad, Ma, Lin, and Zuccon}{Li
  et~al\mbox{.}}{2022}]%
        {li2021improving}
\bibfield{author}{\bibinfo{person}{Hang Li}, \bibinfo{person}{Shengyao Zhuang},
  \bibinfo{person}{Ahmed Mourad}, \bibinfo{person}{Xueguang Ma},
  \bibinfo{person}{Jimmy Lin}, {and} \bibinfo{person}{Guido Zuccon}.}
  \bibinfo{year}{2022}\natexlab{}.
\newblock \showarticletitle{Improving Query Representations for Dense Retrieval
  with Pseudo Relevance Feedback: A Reproducibility Study}. In
  \bibinfo{booktitle}{\emph{Proceedings of the 44rd European Conference on
  Information Retrieval}}.
\newblock


\bibitem[\protect\citeauthoryear{Lin, Ma, Lin, Yang, Pradeep, and Nogueira}{Lin
  et~al\mbox{.}}{2021a}]%
        {lin2021pyserini}
\bibfield{author}{\bibinfo{person}{Jimmy Lin}, \bibinfo{person}{Xueguang Ma},
  \bibinfo{person}{Sheng-Chieh Lin}, \bibinfo{person}{Jheng-Hong Yang},
  \bibinfo{person}{Ronak Pradeep}, {and} \bibinfo{person}{Rodrigo Nogueira}.}
  \bibinfo{year}{2021}\natexlab{a}.
\newblock \showarticletitle{Pyserini: An easy-to-use Python toolkit to support
  replicable IR research with sparse and dense representations}.
\newblock \bibinfo{journal}{\emph{arXiv preprint arXiv:2102.10073}}
  (\bibinfo{year}{2021}).
\newblock


\bibitem[\protect\citeauthoryear{Lin, Yang, and Lin}{Lin
  et~al\mbox{.}}{2021b}]%
        {lin2021batch}
\bibfield{author}{\bibinfo{person}{Sheng-Chieh Lin},
  \bibinfo{person}{Jheng-Hong Yang}, {and} \bibinfo{person}{Jimmy Lin}.}
  \bibinfo{year}{2021}\natexlab{b}.
\newblock \showarticletitle{In-batch negatives for knowledge distillation with
  tightly-coupled teachers for dense retrieval}. In
  \bibinfo{booktitle}{\emph{Proceedings of the 6th Workshop on Representation
  Learning for NLP (RepL4NLP-2021)}}. \bibinfo{pages}{163--173}.
\newblock


\bibitem[\protect\citeauthoryear{Lv and Zhai}{Lv and Zhai}{2009}]%
        {lv2009comparative}
\bibfield{author}{\bibinfo{person}{Yuanhua Lv} {and}
  \bibinfo{person}{ChengXiang Zhai}.} \bibinfo{year}{2009}\natexlab{}.
\newblock \showarticletitle{A Comparative Study of Methods For Estimating Query
  Language Models with Pseudo Feedback}. In
  \bibinfo{booktitle}{\emph{Proceedings of the 18th ACM ACM International
  Conference on Information and Knowledge Management}}.
  \bibinfo{pages}{1895--1898}.
\newblock


\bibitem[\protect\citeauthoryear{Nguyen, Rosenberg, Song, Gao, Tiwary,
  Majumder, and Deng}{Nguyen et~al\mbox{.}}{2016}]%
        {nguyen2016ms}
\bibfield{author}{\bibinfo{person}{Tri Nguyen}, \bibinfo{person}{Mir
  Rosenberg}, \bibinfo{person}{Xia Song}, \bibinfo{person}{Jianfeng Gao},
  \bibinfo{person}{Saurabh Tiwary}, \bibinfo{person}{Rangan Majumder}, {and}
  \bibinfo{person}{Li Deng}.} \bibinfo{year}{2016}\natexlab{}.
\newblock \showarticletitle{MS MARCO: A Human Generated Machine Reading
  Comprehension Dataset}. In \bibinfo{booktitle}{\emph{Workshop on Cognitive
  Computing at NIPS}}.
\newblock


\bibitem[\protect\citeauthoryear{Robertson and Zaragoza}{Robertson and
  Zaragoza}{2009}]%
        {robertson2009probabilistic}
\bibfield{author}{\bibinfo{person}{Stephen Robertson} {and}
  \bibinfo{person}{Hugo Zaragoza}.} \bibinfo{year}{2009}\natexlab{}.
\newblock \showarticletitle{The Probabilistic Relevance Framework: BM25 and
  Beyond}.
\newblock \bibinfo{journal}{\emph{Foundations and Trends in Information
  Retrieval}} \bibinfo{volume}{3}, \bibinfo{number}{4} (\bibinfo{year}{2009}),
  \bibinfo{pages}{333--389}.
\newblock


\bibitem[\protect\citeauthoryear{Rocchio}{Rocchio}{1971}]%
        {rocchio1971rocchio}
\bibfield{author}{\bibinfo{person}{J.J. Rocchio}.}
  \bibinfo{year}{1971}\natexlab{}.
\newblock \showarticletitle{Relevance Feedback in Information Retrieval}. In
  \bibinfo{booktitle}{\emph{The SMART Retrieval System - Experiments in
  Automatic Document Processing}}. \bibinfo{pages}{313--323}.
\newblock


\bibitem[\protect\citeauthoryear{Tsai, Hu, and Chen}{Tsai
  et~al\mbox{.}}{2015}]%
        {tsai2015factors}
\bibfield{author}{\bibinfo{person}{Chih-Fong Tsai}, \bibinfo{person}{Ya-Han
  Hu}, {and} \bibinfo{person}{Zong-Yao Chen}.} \bibinfo{year}{2015}\natexlab{}.
\newblock \showarticletitle{Factors affecting rocchio-based pseudorelevance
  feedback in image retrieval}.
\newblock \bibinfo{journal}{\emph{Journal of the Association for Information
  Science and Technology}} \bibinfo{volume}{66}, \bibinfo{number}{1}
  (\bibinfo{year}{2015}), \bibinfo{pages}{40--57}.
\newblock


\bibitem[\protect\citeauthoryear{Wang, Macdonald, Tonellotto, and Ounis}{Wang
  et~al\mbox{.}}{2021}]%
        {wang2021pseudo}
\bibfield{author}{\bibinfo{person}{Xiao Wang}, \bibinfo{person}{Craig
  Macdonald}, \bibinfo{person}{Nicola Tonellotto}, {and} \bibinfo{person}{Iadh
  Ounis}.} \bibinfo{year}{2021}\natexlab{}.
\newblock \showarticletitle{Pseudo-Relevance Feedback for Multiple
  Representation Dense Retrieval}.
\newblock \bibinfo{journal}{\emph{arXiv preprint arXiv:2106.11251}}
  (\bibinfo{year}{2021}).
\newblock


\bibitem[\protect\citeauthoryear{Xiong, Xiong, Li, Tang, Liu, Bennett, Ahmed,
  and Overwijk}{Xiong et~al\mbox{.}}{2020}]%
        {xiong2020approximate}
\bibfield{author}{\bibinfo{person}{Lee Xiong}, \bibinfo{person}{Chenyan Xiong},
  \bibinfo{person}{Ye Li}, \bibinfo{person}{Kwok-Fung Tang},
  \bibinfo{person}{Jialin Liu}, \bibinfo{person}{Paul Bennett},
  \bibinfo{person}{Junaid Ahmed}, {and} \bibinfo{person}{Arnold Overwijk}.}
  \bibinfo{year}{2020}\natexlab{}.
\newblock \showarticletitle{Approximate Nearest Neighbor Negative Contrastive
  Learning For Dense Text Retrieval}.
\newblock \bibinfo{journal}{\emph{arXiv preprint arXiv:2007.00808}}
  (\bibinfo{year}{2020}).
\newblock


\bibitem[\protect\citeauthoryear{Yang, Fang, and Lin}{Yang
  et~al\mbox{.}}{2018}]%
        {yang2018anserini}
\bibfield{author}{\bibinfo{person}{Peilin Yang}, \bibinfo{person}{Hui Fang},
  {and} \bibinfo{person}{Jimmy Lin}.} \bibinfo{year}{2018}\natexlab{}.
\newblock \showarticletitle{Anserini: Reproducible Ranking Baselines Using
  Lucene}.
\newblock \bibinfo{journal}{\emph{Journal of Data and Information Quality
  (JDIQ)}} \bibinfo{volume}{10}, \bibinfo{number}{4} (\bibinfo{year}{2018}),
  \bibinfo{pages}{1--20}.
\newblock


\bibitem[\protect\citeauthoryear{Yu, Xiong, and Callan}{Yu
  et~al\mbox{.}}{2021}]%
        {yu2021improving}
\bibfield{author}{\bibinfo{person}{HongChien Yu}, \bibinfo{person}{Chenyan
  Xiong}, {and} \bibinfo{person}{Jamie Callan}.}
  \bibinfo{year}{2021}\natexlab{}.
\newblock \showarticletitle{Improving Query Representations for Dense Retrieval
  with Pseudo Relevance Feedback}. In \bibinfo{booktitle}{\emph{Proceedings of
  the 30th ACM International Conference on Information and Knowledge
  Management}}.
\newblock


\bibitem[\protect\citeauthoryear{Zhai and Lafferty}{Zhai and Lafferty}{2001}]%
        {zhai2001model}
\bibfield{author}{\bibinfo{person}{Chengxiang Zhai} {and} \bibinfo{person}{John
  Lafferty}.} \bibinfo{year}{2001}\natexlab{}.
\newblock \showarticletitle{Model-Based Feedback in the Language Modeling
  Approach to Information Retrieval}. In \bibinfo{booktitle}{\emph{Proceedings
  of the 10th ACM International Conference on Information and Knowledge
  Management}}. \bibinfo{pages}{403--410}.
\newblock


\end{thebibliography}

\end{document}